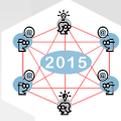

# Invited Abstract: Issues with State-based Energy Consumption Modelling

Torsten Braun, Philipp Hurni

University of Bern
Communication and Distributed Systems
Bern, Switzerland
{braun, hurni}@inf.unibe.ch

Vitor Bernardo, Marilia Curado

University of Coimbra
Center for Informatics and Systems
Coimbra, Portugal
{vmbern, marilia}@dei.uc.pt

*Abstract* — **Energy consumption modelling by state based approaches often assume constant energy consumption values in each state. However, it happens in certain situations that during state transitions or even during a state the energy consumption is not constant and does fluctuate. This paper discusses those issues by presenting some examples from wireless sensor and wireless local area networks for such cases and possible solutions.**

## I. Introduction

Energy consumption is an important issue in wireless networking, in particular in sensor networks with sensor nodes operating for long periods with a single battery as well as in mobile end systems such as smart phones, where users like to enjoy many hours of battery driven operation. Protocols do not only have to support traditional performance metrics such as delay and throughput, but also should support energy-efficient operation. Protocols have to be evaluated in real-world testbeds and simulators. Simulators aim to support accurate energy consumption evaluation by relying on accurate energy models for the wireless device's transceiver, which is often by far the most energy-consuming component in a wireless device.

Typically, energy consumption in simulators is modelled by rather simple state-based approaches, where the time $T_i$ of the transceiver being in a state $i$ is recorded and multiplied with the average current $I_i$ in state $i$ as well as the (often constant) supply voltage $U$ to calculate the consumed energy. We further have to sum up the energy consumed in each state $I_i$ as depicted in Fig. 1. The following formula calculates the overall energy E consumed by the system:

$$E = \sum_{i=1}^{N} T_i \cdot I_i \cdot U$$

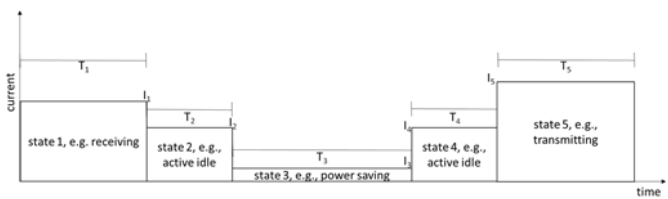

Fig. 1. State-based energy consumption modelling

However, this state-based energy consumption model is only an approximation of the energy consumption in a real system and has some limits in terms of accuracy. As discussed in subsequent sections of this paper the state-based model is somewhat inaccurate, because it is not correct to assume that the current is constant during all states, state transitions do take some time, and the current during the state transition is not equivalent to the current in the previous or subsequent state. Those effects can then lead to some inaccuracy in energy consumption evaluation, as we will discuss using two example scenarios in Sections II and III.

## II. Energy Consumption in Wireless Sensor Networks

In this section, we discuss some energy consumption modelling for wireless sensor nodes [1]. In Fig. 2 we show the measurement of the current flow of a sensor node (MSB430), which is forwarding traffic from a source to a destination node. We identify three states of the node's transceiver, namely sleep, transmitting, and receiving. Fig. 3 shows how the energy consumption of a sensor node is modelled using a so-called three-state-model. Here, the model is not used for simulation but for so-called software-based energy estimation, where the running software of a node measures the time of the transceiver in a state and multiplies the time with the corresponding average current in a state and the supply voltage, similar as discussed in Section I.

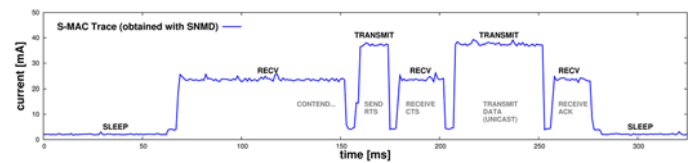

Fig. 2. Current draw of a sensor node [1]

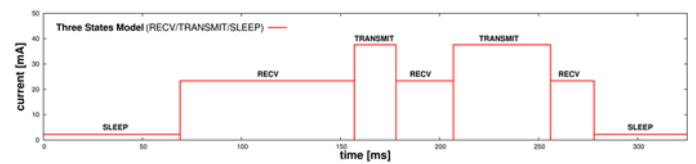

Fig. 3. Sensor node modelled by the three-states model [1]

The software-based energy estimation has been used for the evaluation of several MAC protocols (CSMA, TMAC; SMAC; and WiseMAC) in a wireless sensor network consisting of three nodes, i.e. a source, a forwarder and a destination node, with variable traffic rate. In Fig. 4, dashed and solid lines represent estimated and measured energy consumption, respectively. We can observe that for increasing the traffic rate, measured in packets/s the difference between estimation based on the three-state model and the real evaluation becomes larger. The reason is that for increasing traffic rate the number of state transitions increases too. Comparing Fig. 2 and Fig. 3, we see that in reality there is lower energy consumption during a state transition than what we assume in the software-based estimation using the three-state-model.

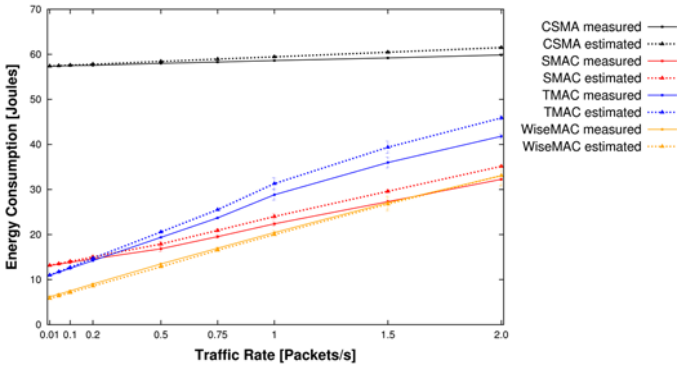

Fig. 4.  Measured vs. Estimated Energy Consumption [1]

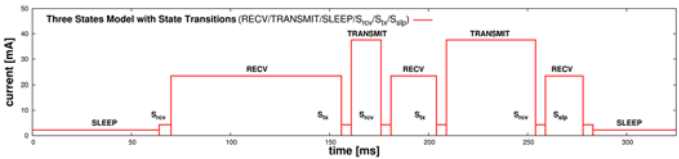

Fig. 5.  Current modelled by the three-states model with state transitions [1]

In order to improve energy consumption estimation, we introduce a better model considering the energy consumption during state transitions, cf. Fig. 5. Fig. 6 shows the measured error of the original three-states model, an improved version of the three-states model using ordinary least square (OLS) regression analysis for estimating energy consumption values of the various states, and the OLS-based three-states model considering state transitions. Fig. 6 shows that by considering state transitions we can improve the estimation accuracy from approximately 4 % to 1 % for the OLS-based three-state model without and with considering state transitions. While the improved model has been applied for software-based energy estimation [1], we could also apply state transitions in a simulation model for more accurate energy consumption modelling.

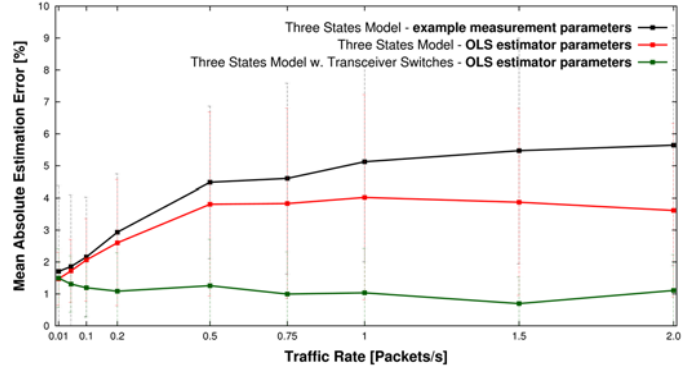

Fig. 6.  Absolute Mean Estimation Error (in %) vs. Traffic Rate (packets/s)[1]

## III. Energy consumption in Wireless Local Area Networks

The IEEE 802.11 standard [3] defines a power management mode that allows the mobile stations turning off both transmitter and receiver capabilities in order to save energy. Fig. 7 shows the simplified state diagram of IEEE 802.11 nodes with power management. Each state consumes a different amount of energy.

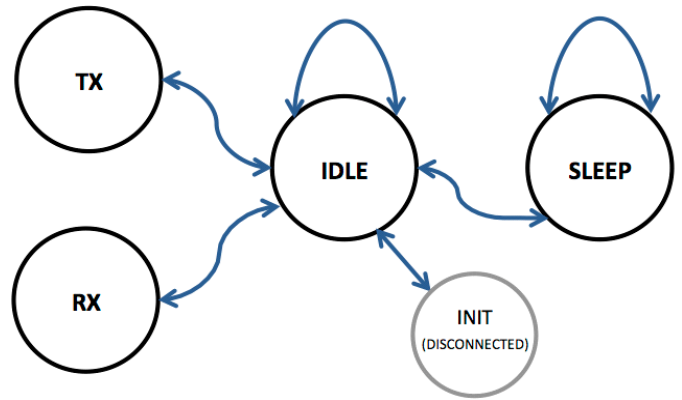

Fig. 7.  Simplified IEEE 802.11 state diagram [2]

Fig. 8 shows the average power consumed (in mW) by an IEEE 802.11 network card (Linksys TP-LINK WN-721n operating in the 2.4GHz frequency band) in three distinct states, namely disconnected, idle and sleep. The depicted values were calculated as average of the 20 runs performance evaluation for each test setup with confidence intervals of 95%. By analysing the error bars for each state, one can observe a lower uncertainty for all the states, showing the suitable accuracy achieved with the employed measurement methodology. See [2] for details about the experiments.

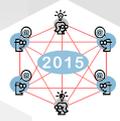

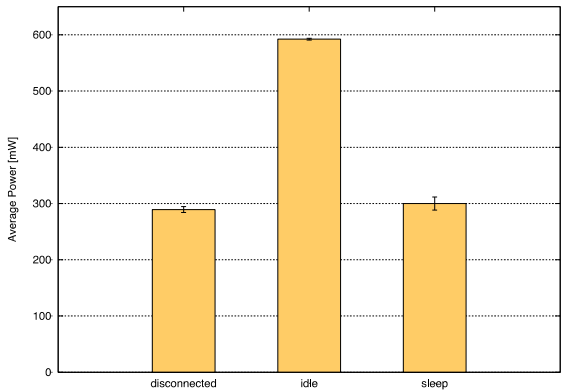

Fig. 8. Average power in disconnected, idle, and sleep states (adapted from [2])

Unlike when operating in disconnected, idle and sleep states, power consumption in the receiving and transmitting states is less constant, as it is depicted in Fig. 9 and Fig. 10. Fig. 9 shows the measured power consumption of the same IEEE 802.11 network card without enabled power management. Fig. 10 shows the power consumption of the same network card with enabled power management. Fig. 9 and Fig. 10 show strong fluctuations during sending/receiving dependent on the network traffic. Moreover, there is some phase (connecting), where the node is moving from disconnected towards connected state. We can compare this phase to a state transition as discussed in Section II. Again, we see that the energy consumption during this phase/state transition is somewhat different from the previous and subsequent state. In this case, energy consumption during state transition is somewhat lower (due to negative peaks) than in the connected state, but significantly higher when compared to the disconnected state.

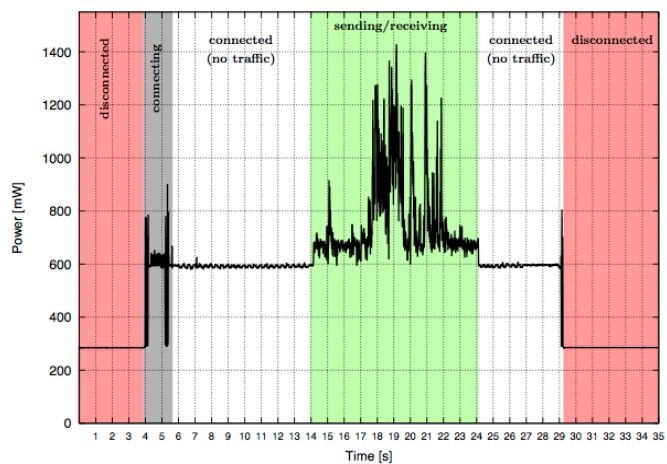

Fig. 9. IEEE 802.11 network card power consumption without power management [2]

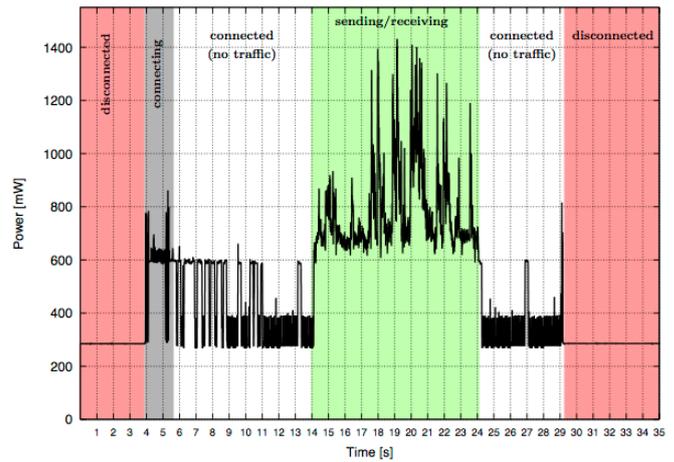

Fig. 10. IEEE 802.11 network card power consumption with power management [2]

Fig. 11 zooms four state transitions of the power consumption measurements from Fig. 10, namely connecting (Fig. 11a), starting transmission/reception after being connected without traffic (Fig. 11b), stopping transmission/ reception before being connected without traffic (Fig. 11c), and disconnecting (Fig. 11d). In Fig. 11 b-d, we see some periodic peaks while being connected without traffic. This is further zoomed in Fig. 11c. Those peaks are caused by receiving beacon frames from the access point. In this scenario beacons are configured to be sent by the access point every 100 ms. The peaks increase the average energy consumed and such effects have to be considered for accurate energy modelling.

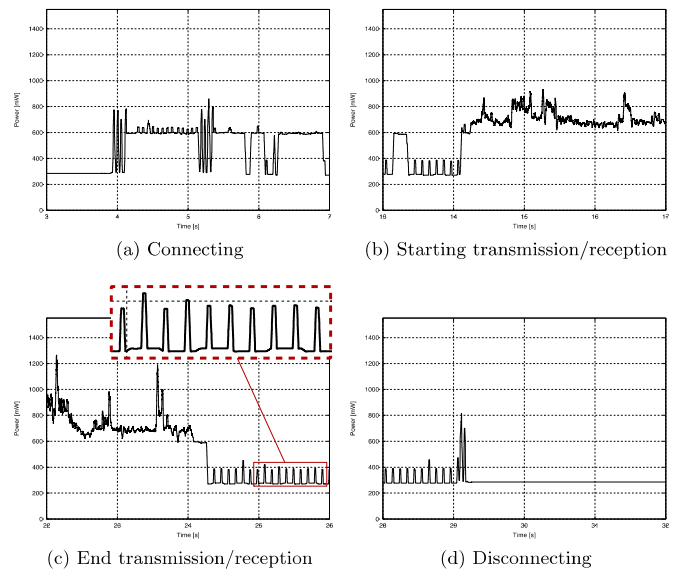

Fig. 11. Detailed states transition power consumption pattern with power saving enabled [2]

## IV. DISCUSSIONS AND CONCLUSIONS

In this paper, we have presented energy measurement results from previous work on energy consumption evaluations in wireless sensor and wireless local area networks. We have made the following observations:

1. Energy consumption during state transitions can significantly differ from previous and subsequent states.

2. During an active state, e.g., transmitting, receiving, active idle/connected without traffic, the energy consumption can vary dependent on the current traffic transmitted or received. This also includes reception of control messages, as it is the case for IEEE 802.11 beacons.

For accurate evaluation of energy consumption in either software-based energy estimation or simulation, where state-based energy consumption models have been applied in the past, we might need to more accurately model state transitions and dynamic fluctuations within a state. In particular, in [1] we have shown that we can significantly decrease the estimation error by considering the behaviour during a state transition.

Further improvements might result from considering other parameters such as the number and size of received/transmitted data/control messages during a state. More measurements of specific wireless network hardware are needed to investigate the impact of such parameters to energy consumption.

### REFERENCES

[1] Philipp Hurni, Benjamin Nyffenegger, Torsten Braun, Anton Hergenroeder: On the Accuracy of Software-Based Energy Estimation Techniques, Wireless Sensor Networks, Lecture Notes in Computer Science Volume 6567, Springer-Verlag, 2011, pp. 49-64

[2] Vitor Bernardo, Marilia Curado, Torsten Braun: An Overview of Energy Consumption in IEEE 802.11 Access Networks, in Wireless Networking for Moving Objects, A. Kassler, I. Ganchev, M. Curado (eds.), Lecture Notes in Computer Science Volume 8611, Springer-Verlag, 2014, pp. 157-176

[3] IEEE 802.11 Standards Association: IEEE Standard for Information technology - Telecommunications and information exchange between systems, Local and metropolitan area networks - Specific requirements, Part 11: Wireless LAN Medium Access Control (MAC) and Physical Layer (PHY) Specifications, IEEE Std. 802.11-2012 (Revision of IEEE Std 802.11-2007)